\documentclass[a4paper,11pt]{article}

\usepackage{jheppub}
\usepackage{multirow}
\usepackage{graphicx}
\usepackage{dcolumn}
\usepackage{bm}
\usepackage{amsmath}
\usepackage{amssymb}
\usepackage{mathrsfs}
\usepackage{amsfonts}
\usepackage{color, soul}
\pdfoutput=1

\title{Testing the light dark matter scenario of the MSSM at the LHC}

\author[a,b]{Junjie Cao,}
\author[a]{Yangle He,}
\author[a]{Liangliang Shang,}
\author[c]{Wei Su,}
\author[c]{Yang Zhang}

\affiliation[a]{Department of Physics,
                Henan Normal University, Xinxiang 453007, China}
\affiliation[b]{Department of Applied Physics, Xi'an Jiaotong University, Xi'an 710049, China}
\affiliation[c]{State Key Laboratory of Theoretical Physics, Institute of Theoretical Physics,
                Academia Sinica, Beijing 100190, China}

\emailAdd{junjiec@itp.ac.cn}
\emailAdd{heyangle90@gmail.com}
\emailAdd{shlwell1988@gmail.com}
\emailAdd{weisv@itp.ac.cn}
\emailAdd{zhangyang@itp.ac.cn}

\abstract{In the light dark matter (DM) scenario of the MSSM, the DM relic density puts non-trivial requirements on the spectrum of supersymmetric particles. As a result,
the direct search for multi-lepton signals at the LHC has great impact on the scenario. In this work, we concentrate on the searches for
sleptons and electroweak-inos at the LHC, and investigate the constraints on the light DM scenario from the 8 TeV LHC data as well as the capability of the 14 TeV LHC to test the scenario.
We first get the samples of the scenario by scanning the vast parameter space of the MSSM and considering some easily available  constraints, such as those from the DM relic
density, the LUX experiment and the Higgs searches at colliders.  Then for the surviving samples, we simulate the $2l+E_T^{miss}$ signal from slepton pair production process
and the  $2l+E_T^{miss}$ and $3l+E_T^{miss}$ signals from chargino and neutralino associated production processes at both the 8 TeV LHC and the 14 TeV LHC. Our simulations indicate that
the 8 TeV LHC has excluded a sizable portion of the samples, and the 14 TeV LHC can be even much more powerful in testing the scenario. For example, in case that no multi-lepton signals
are observed at the 14 TeV LHC, most samples of the light DM scenario will be excluded, especially a lower limit on the lightest neutralino mass
will be set at 42 GeV and 44 GeV with 30 $fb^{-1}$ and  100 $fb^{-1}$ data respectively, and this limit can be further pushed up to $55\  {\rm GeV}$  with
300 $fb^{-1}$ data.}
\begin{document}
\maketitle
\newpage

\section{Introduction}

The Minimal Supersymmetric Standard Model (MSSM) is one of the most promising new physics model beyond the Standard Model (SM), which can
stabilize the electroweak (EW) scale, explain the cosmic dark matter (DM) and achieve the gauge coupling unification simultaneously\cite{MSSM-1,MSSM-2,MSSM-3}.
In recent years,  a large number of searches for the supersymmetric particles (sparticles) predicted by the MSSM
have been  carried out at the LHC experiments, and consequently, stronger limits on the spectrum of the sparticles than the LEP experiments have been obtained.
For example, the null results in the searches for multi-jets plus large missing transverse energy ($E_T^{miss}$) signal have set the lower mass bounds for colored sparticles at TeV scale,
i.e. about $1.2 {\rm TeV}$ and $0.8 {\rm TeV}$ for gluino and degenerate first two generation squarks
respectively in optimal case\cite{ATLAS-Multi-jets,CMS-Multi-jets}; and as for the EW sparticles of the MSSM, although they are less
constrained due to their relatively small direct production rate, the limits can still be up to $300 {\rm GeV}$ for the sleptons\cite{ATLAS-Sletpon,CMS-Sletpon} and up to
$700 {\rm GeV}$ for the charginos and neutralinos\cite{ATLAS-Chargino,CMS-Sletpon}.  More strikingly, with the recent operation of the LHC Run-II at $13\  {\rm TeV}$ of center of mass energy,
it is widely expected that much heavier sparticles will be explored very soon, which will be helpful to improve our understanding on
some fundamental questions about nature. Obviously, discussing the potential of the LHC experiments to test the MSSM is an important task 
for both theorists and experimentalists, and such studies have been intensively carried out (e.g. in the very recent papers \cite{EW-search1,EW-search2,EW-search3,EW-search4,EW-search5}, the searches for the EW sparticles were investigated comprehensively).   

In the MSSM with R-parity, the lightest neutralino is usually the lightest sparticle, and thus can act as the DM candidate\cite{MSSM-3}. So far the scenario
featured by a moderately light DM has been intensively studied\cite{LDM-1,LDM-2,LDM-3,LDM-4,LDM-5,LDM-6,LDM-7,LDM-8,LDM-9,LDM-10,LDM-11,LDM-12,LDM-13,LDM-14,LDM-15,
LDM-16,LDM-17,LDM-18,LDM-19,LDM-20,LDM-21,LDM-22,LDM-23,LDM-24,Cao-LDM,LDM-25,LDM-26,LDM-27,LDM-28,LDM-29,LDM-30}. One motivation for this is that in some fundamental theories
such as the minimal supergravity theories\cite{mSUGRA}, the EW sparticles tend to be significantly lighter than the colored sparticles. This pattern of the sparticle spectrum does not conflict with any constraints from low energy processes as well as from the direct searches for the sparticles at the colliders, and instead it is helpful to solve some
experimental anomalies such as the discrepancy of the measured muon anomalous magnetic moment from its SM prediction and the Galactic Center $\gamma$-ray excess observed by the Fermi-LAT\cite{GCE-1,GCE-2,GCE-3,GCE-4}. Another motivation for the scenario is that light higgsinos are the minimal tree-level requirement posed by naturalness. However, a light higgsino-like DM can not account for the full amount of the observed relic density of the DM since it annihilated too efficiently in early universe\cite{DM-annihilation-1}.
Consequently, simultaneous presence of light bino and higgsinos, which will mix to form a light DM, is the minimal ingredient of a natural MSSM\cite{DM-annihilation-2}.
In this work, we are particularly interested in the DM lighter than about $100 {\rm GeV}$. In this case, the chargino mass limit from the LEP experiments has required the DM to be bino-dominated. Then the weak interaction of the DM together with the sizable mass splitting of the DM from the other sparticles typically leads to the overproduction of the DM in the early universe, unless that an efficient annihilation mechanism was at work\cite{LDM-26,LDM-27}. This situation sets non-trivial requirement on the spectrum of the sparticles, especially given that the spectrum has been limited by the direct search for SUSY at colliders. As a result, only a small corner of the MSSM parameter space is relevant to our discussion, which makes it possible to test the scenario by certain signals of the sparticles at the LHC\cite{LDM-26,LDM-27}.

Recent discussions on the light DM scenario concentrated on the complement of new experimental constraints, such as the 125 GeV Higgs data and the direct searches for sparticles
at the LHC, and as a result, the allowed parameter space of the scenario becomes more and more constrained\cite{LDM-20,LDM-21,LDM-22,LDM-23,LDM-24,Cao-LDM,LDM-25,LDM-26,LDM-27}.
One impressive result is that, after considering the LHC searches for multi-$\tau$ plus large $E_T^{miss}$ signal, the DM lighter than about $30 {\rm GeV}$ has been
excluded in the MSSM \cite{LDM-20}. In this context, we extend the latest relevant analysis in this field \cite{LDM-27} by relaxing its assumptions
on the parameters of the MSSM, and more important, we focus on the parameter space surviving current
experimental constraints to investigate the capability of the LHC experiments in testing the light DM scenario.
As we will show below,
the light DM annihilated mainly via s-channel exchange of a $Z$ boson or SM-like Higgs boson to satisfy the bound from the relic density, and in such a process, the corresponding resonant enhancement played a crucial role. While on the other hand, the effective coupling of the DM with nuclei in such a case usually drops drastically, and so is the rate of the DM annihilation in Galactic Center at present day, which is relevant for indirect DM searches. Hence our study is a useful supplement to DM direct and indirect detection experiments in exploring the light DM scenario. Moreover, we note that for most LHC searches for sparticles, they rely heavily on large $E_T^{miss}$ signal contributed by DM. So deciphering the property of the DM is important for the searches. As we are going to show, if nature has chosen the specific scenario of the MSSM, the future LHC experiment is powerful to test it, especially DM lighter than about $55\ GeV$ is disfavored, and without the presence of light sleptons, the mass of the DM is fixed at about one half of the SM-like Higgs boson mass if no SUSY signal is detected at the 14 TeV LHC.

The rest of this paper is organized as follows. In section 2 we review the features of the electroweak-inos in the MSSM, and point out that the DM relic density can impose non-trivial constraints
on the higgsino mass $\mu$ in light DM scenario. In section 3 we scan the parameter space of the MSSM by considering some easily available constraints to obtain the samples of the scenario. Then we pay special attention to the important constraints from the direct searches for $2 l + E_T^{miss}$ and $3 l + E_T^{miss}$ signals at the 8 TeV LHC by detailed simulation,
and check whether the samples can survive them
in section 4. In section 5 we extend the simulation study to the 14 TeV LHC and discuss its potential to prob the scenario. As a useful supplement to the direct searches, we also briefly examine the capabilities of the future DM direction experiments to detect the scenario in section 6. Finally, we draw our conclusions in section 7.

\section{The electroweak-inos in the MSSM}

In the MSSM, the fields bino $\tilde{B}^0$, wino $\tilde{W}^0$, and higgsinos $\tilde{H}_{d}^0$ and $\tilde{H}_{u}^0$ will mix to form mass eigenstates, which are usually called
as neutralinos $\tilde{\chi}_i^0$ ($i=1,\cdots 4$).  In the basis $(\tilde{B}^0, \tilde{W}^ 0, \tilde{H}_d^0, \tilde{H}_u^0)$, the mass matrix of the fields is given by
\begin{eqnarray}
{\cal{M}}_{\tilde{\chi}^0} = \left( \begin{array}{cccc}
  M_1 			      & 0 				      & -m_Z s_W c_\beta & m_Z s_W s_\beta \\
  0				      & M_2 			      & m_Z c_W c_\beta  & -m_Z c_W s_\beta \\
  -m_Z s_W c_\beta & m_Z c_W c_\beta  & 0 				      & -\mu 			      \\
  m_Z s_W s_\beta  & -m_Z c_W s_\beta &  -\mu 			      & 0				      \\
   \end{array} \right),
\end{eqnarray}
where $M_1$ and $M_2$ are the soft bino and wino masses respectively, $\mu$ represents the higgsino mass, $c_\beta = \cos\beta$ and
$s_\beta =\sin\beta$ with $\tan \beta\equiv v_u/v_d$ being the ratio of the vacuum expectation values of the two Higgs doublet.
This mass matrix can be diagonalized by an unitary $4 \times 4 $ matrix $N$ so that the interactions of the neutralinos are given by
\begin{eqnarray} \label{interaction}
{\cal{L}}_{\tilde{\chi}^0} & = & \tilde{l}_L^\ast \bar{\tilde{\chi}}_i^0 \Big[ {{e} \over {\sqrt{2}s_w c_w}}( N_{i1}s_w
+ N_{i2}c_w ) P_L  + y_l N_{i3}^\ast P_R \Big] l \nonumber \\
& &  +  \tilde{l}_R^\ast {\bar{\tilde{\chi}}}_i^0 \Big[ {{- \sqrt{2}e}
\over {c_w}} N_{i1}^\ast P_R +y_l N_{i3} P_L \Big] l +  \tilde{\nu}^\ast \bar{\tilde{\chi}}_i^0 \Big[ {{e} \over {\sqrt{2}s_w c_w}}( N_{i1}s_w
- N_{i2}c_w ) P_L   \Big] \nu  \nonumber \\
& & +  \frac{e}{2 s_w} h \bar{\tilde{\chi}}_i^0
\Big[ (N_{i2} - N_{i1} \tan \theta_w ) (\sin \alpha N_{j3} + \cos \alpha N_{j4} ) + (i \leftrightarrow j ) \Big]{\tilde{\chi}}_j^0  \nonumber \\
&& +  \  {e \over {s_w c_w}}Z_\mu \bar{\tilde{\chi}}_i^0 \gamma^\mu({\cal{O}}_{ij}^L P_L + {\cal{O}}_{ij}^R P_R){\tilde{\chi}}_j^0 + \cdots,
\end{eqnarray}
where $y_l$ is the Yukawa coupling coefficient for the lepton $l$, $h$ denotes the SM-like Higgs boson, $c_W = \cos \theta_W$, $s_W = \sin \theta_W$ and
${\cal{O}}_{ij}^L= - {\cal{O}}_{ij}^{R \ast} = {-}{1 \over 2}N_{i3}N_{j3}^\ast + {1 \over 2}N_{i4}N_{j4}^\ast $.
The Lagrangian in Eq.(\ref{interaction}) indicates that the $Z \bar{\tilde{\chi}}_i^0 \tilde{\chi}_j^0$ interaction is determined by the higgsino components of the neutralinos,
and by contrast the $h \bar{\tilde{\chi}}_i^0 \tilde{\chi}_j^0$ interaction is determined by the gaugino component of one of the neutralinos, and also the higgsino
component of the other neutralino.

Assuming $M_1 < |\mu| \ll M_2$, one can expand the matrix $N$ by powers of $M_1/\mu$, and up to the first order of the expansion, the matrix is given by\cite{LDM-27}
\begin{eqnarray} \label{matrix-approximation}
N \simeq \left( \begin{array}{cccc}
  1 & 0 & \frac{m_Zs_W}{\mu}(s_\beta+c_\beta\frac{M_1}{\mu}) & -\frac{m_Zs_W}{\mu}(c_\beta+s_\beta\frac{M_1}{\mu}) \\
  \frac{m_Zs_W(s_\beta+c_\beta)}{\sqrt{2}\mu}(1+\frac{M_1}{\mu}) & 0 & -\frac{1}{\sqrt{2}} & \frac{1}{\sqrt{2}} 			 \\
  \frac{m_Zs_W(s_\beta- c_\beta)}{\sqrt{2}\mu}(1- \frac{M_1}{\mu}) & 0 & -\frac{1}{\sqrt{2}} & -\frac{1}{\sqrt{2}}			 \\
  0															     & -1& 0 				  & 0 						\\
   \end{array} \right).
\end{eqnarray}
Then for a bino-like $\tilde{\chi}_1^0$ and a higgsino-like $\tilde{\chi}_k^0$,  one can conclude from Eq.(\ref{interaction}) and Eq.(\ref{matrix-approximation}) that
\begin{eqnarray} \label{couplings}
C_{\tilde{l}^\ast \bar{\tilde{\chi}}_1^0 l} &\propto& e, \quad C_{\tilde{l}^\ast \bar{\tilde{\chi}}_k^0 l} \propto y_l, \quad C_{\tilde{\nu}^\ast \bar{\tilde{\chi}}_1^0 \nu} \propto e, \quad C_{\tilde{\nu}^\ast \bar{\tilde{\chi}}_k^0 \nu} \sim 0, \nonumber \\
C_{h \bar{\tilde{\chi}}_1^0 \tilde{\chi}_1^0} &\propto& e \frac{m_Z}{\mu} \Big[ \cos (\beta + \alpha) + \sin (\beta -\alpha) \frac{M_1}{\mu} \Big], \quad C_{h \bar{\tilde{\chi}}_1^0 \tilde{\chi}_k^0} \propto e ( \sin \alpha \pm \cos \alpha ), \nonumber \\
C_{Z \bar{\tilde{\chi}}_1^0 \tilde{\chi}_1^0} &\propto& e \frac{m_Z^2}{\mu^2} \cos 2 \beta ( 1 - \frac{M_1^2}{\mu^2} ), \quad
C_{Z \bar{\tilde{\chi}}_1^0 \tilde{\chi}_k^0} \propto e \frac{M_Z}{\mu} ( s_\beta \pm c_\beta) (1 \pm \frac{M_1}{\mu} ),
\end{eqnarray}
where $C_{XYZ}$ represents the coupling coefficient for the interaction involving the particles $X$, $Y$ and $Z$. Eq.(\ref{couplings}) indicates that if the $\tilde{\chi}_1^0$ as the light DM candidate
annihilated in early universe mainly by $s$-channel exchange of a $Z$ boson or the SM-like Higgs boson\cite{LDM-26,LDM-27}, an upper bound on $\mu$ has to be imposed to forbid
its overproduction, and similarly an upper bound on slepton mass can be obtained if the DM annihilation proceeded mainly by $t/u$-channel slepton
mediation. On the other side, noting that neutralinos and charginos as well as sleptons are going to be intensively explored at the 14 TeV LHC,
we expect that the light DM scenario considered in this work will be readily tested at this machine. So it is necessary to discuss the potential of the LHC in this respect. This is
the main task of this work.

In this work, we focus on the parameter space of the MSSM where the $s$-channel annihilations play the dominant role,
but noting that current bound on slepton masses is rather weak, we also allow for the presence of light sleptons. Obviously, in the case that the sleptons
contribute significantly to the annihilations, the bound on $\mu$ will be relaxed greatly. Another impact of the light sleptons is that they may affect the decay
of the neutralinos, i.e. in addition to the decays $\tilde{\chi}_k^0 \to Z \tilde{\chi}_1^0, h \tilde{\chi}_1^0$, the decay mode
$ \tilde{\chi}_k^0 \to \tilde{l}^\ast l \to \bar{l} l \tilde{\chi}_1^0$ may be open, and consequently the LHC search for the neutralinos becomes
quite complicated.

\section{Light DM scenario in the MSSM}

In our study we get the light DM scenario by scanning the parameter space of the MSSM. To simplify the analysis, we make following assumptions
about the involved parameters:
\begin{itemize}
\item The masses of gluino and the first two generation squarks are fixed at 2 TeV, which are above their mass
limits set by the LHC searches for SUSY.
\item With regard to the third generation squarks, we assume $m_{U_3}=m_{D_3}$ for the right-handed soft
breaking masses and $A_t = A_b$ for soft breaking trilinear coefficients, and let the other parameters free to tune the SM-like Higgs boson mass.
\item We take a common value $m_{\tilde{l}}$ for all soft parameters in slepton sector, i.e. $m_{L_{1,2,3}}=m_{E_{1,2,3}}=A_{E_{1,2,3}} \equiv m_{\tilde{l}}$,
and treat $m_{\tilde{l}}$ as a free parameter since we note that light sleptons can play a role in the DM annihilation.
\end{itemize}
As shown in previous studies\cite{LDM-1,LDM-2,LDM-3,LDM-4,LDM-5,LDM-6,LDM-7,LDM-8,LDM-9,LDM-10,LDM-11,LDM-12,LDM-13,LDM-14,LDM-15,
LDM-16,LDM-17,LDM-18,LDM-19,LDM-20,LDM-21,LDM-22,LDM-23,LDM-24,Cao-LDM,LDM-25,LDM-26,LDM-27}, these assumptions do not affect
the features of the light DM scenario.

\begin{figure}[t]
  \centering
  \includegraphics[width=15cm]{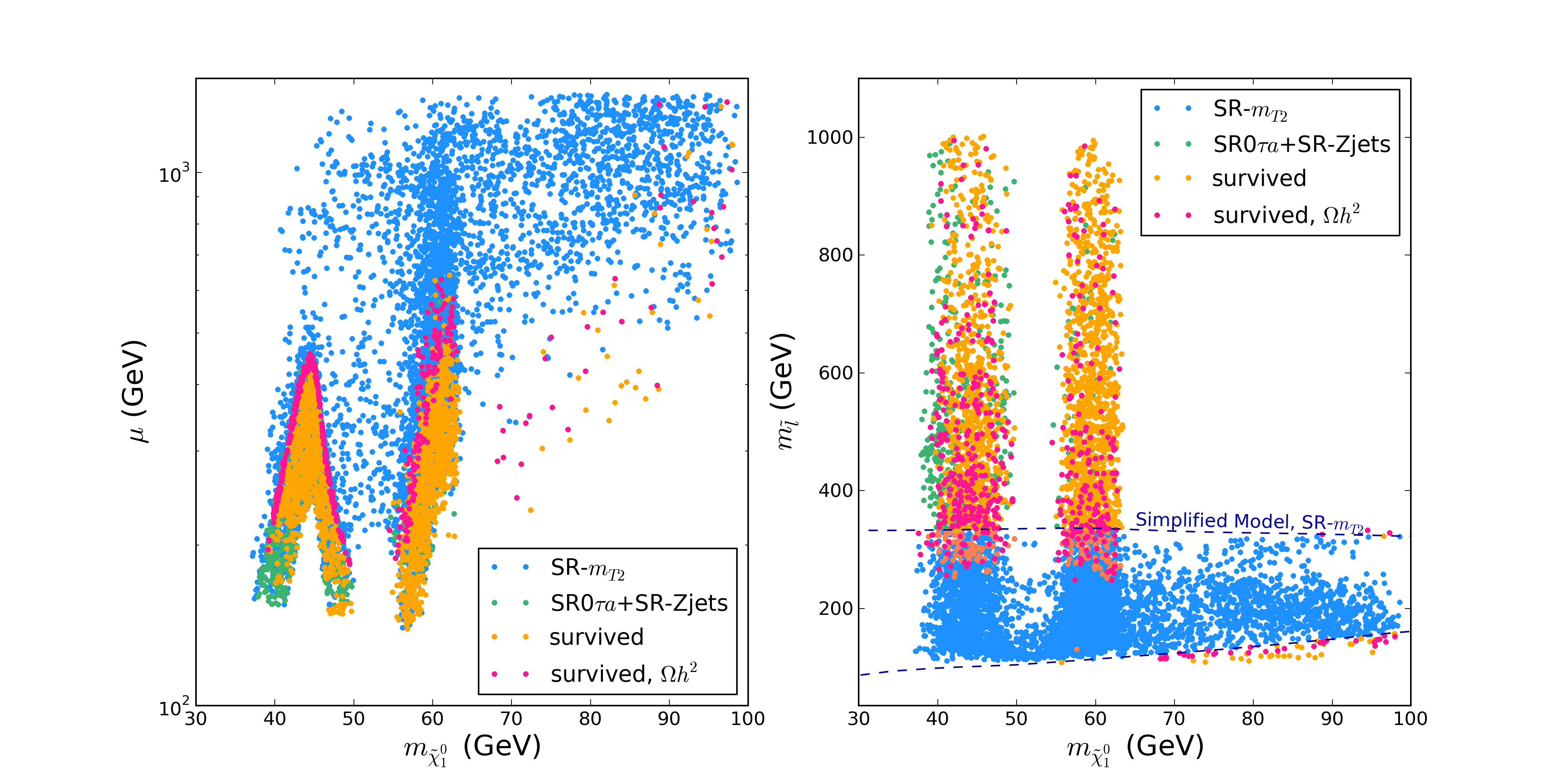}
  \caption{Samples surviving the constraints considered in section 3, which are projected on the $\mu-m_{\tilde{\chi}_1^0}$ and $m_{\tilde{l}}-m_{\tilde{\chi}_1^0}$ planes.
  The blue ones are further excluded by the signal region of SR-$m_{T2}$ in the direct search for sleptons with $2l+E_T^{miss}$ signal at the $8 \ {\rm TeV}$ LHC,
   and the green ones are further excluded by the combination of the signal regions SR0${\tau a}$ and SR-Z jets in the search for $3l+E_T^{miss}$ and $2l+E_T^{miss}$
   signals from the charginos and neutralinos associated production processes at the $8\  {\rm TeV}$ LHC. The red and orange ones are the remaining samples
   with the former able to get the measured DM relic density at $3 \sigma $ level and the latter only satisfying the $3 \sigma$ upper bound of the density.}
  \label{fig1}
\end{figure}

Now the free parameters include $tan\beta$, $M_1$, $M_2$, $\mu$, $m_A$, $m_{\tilde{l}}$, $M_{Q_3}$, $M_{U_3}$ and $A_t$. We define
all these parameters except for $\tan \beta$ at the scale of 2 TeV, and scan the following parameter space\footnote{We note that in the limit $\lambda, \kappa \to 0$ of the Next-to-Minimal Supersymmetric Standard Model (NMSSM), the phenomenology of the NMSSM is same as that of the MSSM for fixed value of $\mu$. So we use the multipurpose package NMSSMTools \cite{NMSSMTools} to perform the scan. As pointed out by the authors of the package, it can reproduce correctly the results of the MSSM.} :
\begin{eqnarray}
&& ~ 2< \tan\beta < 60, ~10~{\rm GeV}<M_1<100~{\rm GeV}, 100~{\rm GeV} < M_2 < 1000~{\rm GeV},  \nonumber\\
&&  ~ 100~{\rm GeV}< \mu < 1500 ~{\rm GeV}, ~50 ~{\rm GeV} < M_A <2 ~{\rm TeV}, \nonumber\\
&&  ~ |A_t| < 5 ~{\rm TeV}, ~ 200 ~{\rm GeV} < m_{Q_3}, m_{U_3} < 2 ~{\rm TeV}, ~100 ~{\rm GeV} < m_{\tilde{l}} < 2 ~{\rm TeV}.
\end{eqnarray}
In the scan, we consider following easily available constraints:
\begin{itemize}
  \item Firstly, we impose the constraints from the LEP searches for SUSY, which include the lower mass limits of charginos and sleptons, $m_{\tilde{\chi}_i^{\pm}}>103.5$ GeV and $m_{\tilde{l}}>93.2$ GeV, the upper bounds on the cross section $\sigma (e^+ e^- \to \tilde{\chi}_1^0 \tilde{\chi}_i^0) \lesssim 0.05 {\rm pb}$ for $i > 1$ and the non-SM invisible decay width of Z boson $\Gamma_{Z\to\tilde{\chi}_1^0\tilde{\chi}_1^0} \leq 1.71$ MeV.
  \item Secondly, we consider the constraints from B-physics, such as the precise measurements of $B\to X_s \gamma$, $B_s\to \mu^+\mu^-$, $B_d \to X_s \mu^+ \mu^-$
      and the mass differences $\Delta M_d$ and $\Delta M_s$ at 2$\sigma$ C.L.\cite{PDG}.
  \item Thirdly, we require the samples to explain the discrepancy of the measured value of the muon anomalous magnetic moment from its SM prediction at 2$\sigma$ level, i.e. $12.7 \leq  \delta a_{\mu}^{SUSY} \leq 44.7$\cite{PDG}.
  \item Fourthly, we implement the constraints on the Higgs sector of the MSSM with the packages HiggsBounds\cite{HiggsBounds} and HiggsSignal\cite{HiggsSignals}, including the
      fit to the $125 GeV$ Higgs data collected at the LHC.
  \item Fifthly, we require the bino-dominated $\tilde{\chi}_1^0$ to take up more than $10\%$ component of the total DM, and meanwhile its relic density smaller than the 3$\sigma$ upper limit of the PLANCK\cite{relic:Planck} and WMAP 9-year data\cite{relic:WMAP}, i.e. $\Omega h^2 \leqslant 0.131$ where a $10\%$ theoretical uncertainty is included. We also impose the LUX
      exclusion bound on DM-nucleon scattering cross section at $90\%$ C.L.\cite{direct:LUX}. In the case that the $\tilde{\chi}_1^0$ is only a fraction $\epsilon$ of the total DM,
      we assume that the other components of the DM have no interaction with nucleon, and consequently, we can implement the constraint of the LUX experiment by scaling the experimental upper bound of the cross section with a factor $1/\epsilon$.  In our analysis, both the relic density and the scattering rate are obtained by the code micrOMEGAs\cite{micrOMEGA}.
  \item Finally, we impose constraints from the LHC searches for third generation squarks by the code FastLim\cite{Fastlim}. This code contains the results of various experimental analyses in searching for third generation squarks, and thus provides a fast way to implement the constraints.
\end{itemize}

The samples surviving above constraints are projected on the $\mu-m_{\tilde{\chi}_1^0}$ and $m_{\tilde{l}}-m_{\tilde{\chi}_1^0}$ planes in Fig.\ref{fig1} (we will explain the meanings of the samples marked by different colors later). This figure shows that the bino-like $\tilde{\chi}_1^0$ must be heavier than about $37 {\rm GeV}$, and a large portion of the samples are centered around
$m_{\tilde{\chi}_1^0} \simeq m_Z/2$ or $m_{\tilde{\chi}_1^0} \simeq m_h/2$. Beside these, we find that the surviving samples can be classified into following three types:
\begin{itemize}
\item Type-I samples: those featured by $m_{\tilde{l}} \lesssim 350 {\rm GeV}$. For this type of samples, we checked that the $t/u$-channel mediation of the sleptons played an important role in the DM annihilation in early universe.
\item Type-II samples: those featured by $m_{\tilde{l}} \gtrsim 350 {\rm GeV}$ and $m_{\tilde{\chi}_1^0} \simeq m_Z/2$.  This type of samples annihilated mainly by $s$-channel exchange of a $Z$ boson, and consequently, as we discussed in last section the DM relic density requires $\mu \lesssim 470 {\rm GeV}$.
\item Type-III samples: those featured by $m_{\tilde{l}} \gtrsim 350 {\rm GeV}$ and $m_{\tilde{\chi}_1^0} \simeq m_h/2$.  This type of samples annihilated mainly by $s$-channel exchange of the SM-like Higgs boson, and the density requires $\mu \lesssim 800 {\rm GeV}$.
\end{itemize}
Moreover, we checked that the stops in the surviving samples must be heavier than about $300 {\rm GeV}$, and for the samples with $m_{\tilde{t}_1} \simeq 300 {\rm GeV}$, the $\tilde{t}_1$ mainly decays into higgsino-dominated neutralinos or chargino as the first step, and the higgsinos subsequently decay into the $\tilde{\chi}_1^0$. Due to the lengthened decay chain, the constraint from the direct searches for stops at the LHC is weakened.

\section{Constraints from the multi-lepton signals at the 8 TeV LHC}

From Fig.\ref{fig1}, one can learn that most of the surviving samples are characterized by predicting either moderately light sleptons or moderately low $\mu$.
This motivates us to further constrain the samples by the direct searches for sleptons and neutralinos/charginos at the $8 \ {\rm TeV}$ LHC by detailed simulation
\footnote{We note that, since the $\tilde{\chi}_1^0$ in our scenario is bino dominated and $m_{\tilde{\chi}_1^0} \gtrsim 37 {\rm GeV}$, the constraint from the mono-jet search presented in \cite{monojet} should be very weak since the production rate for the process $p p \to \bar{\tilde{\chi}}_1^0 \tilde{\chi}_1^0 j$ is small.}. In the rest of this section, we
consider following experimental analyses:

\begin{table}[t]
\small
\centering
\caption{The selections of $N_{j}$, $\Delta m_{ll,Z}$, $E_{T,rel}^{miss}$, $P_{T}^{ll}$, $m_{T2}$,  $\Delta R_{ll}$ and $m_{jj}$ for different SRs of the SR-$m_{T2}$ and
the SR-Zjets.  The expected cross sections of the SM backgrounds for each SR at the 14 TeV LHC are also presented,
which will be used later.
Quantities with mass dimension and the cross sections are given in units of GeV and fb respectively.}
\label{table1}
\begin{tabular}{c|ccccccc|cccc}
\hline
SR                 & $N_j$ & $\Delta m_{ll,Z}$ & $E_{T,rel}^{miss}$ & $P_{T}^{ll}$ & $m_{T2}$ & $\Delta R_{ll}$ & $m_{jj}$ & WW   & ZV   & Other & Total \\
\hline
$m_{T2}^{90}$   & 0            & $>10$ & $-$     & $-$      & $>90$     & $-$         & $-$        & 1.71 & 1.36 & 0.26  & 3.33  \\
$m_{T2}^{120}$ & 0            & $>10$ & $-$     & $-$      & $>120$   & $-$         & $-$        & 0.12 & 0.44 & 0.00  & 0.57  \\
$m_{T2}^{150}$ & 0            & $>10$ & $-$     & $-$      & $>150$   & $-$         & $-$        & 0.02 & 0.19 & 0.00  & 0.21  \\
\hline
Zjets                   & $\geq 2$ & $<10$ & $>$80 & $>$80 & $-$         & [0.3,1.5] & [50,100] & 0.02 & 0.14 & 0.03  & 0.19 \\
\hline
\end{tabular}
\end{table}

\begin{itemize}
\item The search for $2 l + E_T^{miss}$ signal from slepton pair production process or electroweak-ino pair production process
at the 8 TeV LHC with $20.3 \ {\rm fb}^{-1}$ integrated luminosity\cite{2lETmiss}.  In this analysis, seven signal regions (SRs) were defined. The first three, collectively
referred to as SR-m$_{T2}$, were designed to provide sensitivity to the process
$ p p \to \tilde{l}^\ast \tilde{l} \to  2 l + E_T^{miss}$. The next three,
SR-WW, were designed to be sensitive to $p p \to \bar{\tilde{\chi}}_i^\pm \tilde{\chi}_j^\pm \to
(\tilde{\chi}_1^0 W^\mp) (\tilde{\chi}_1^0 W^\pm ) \to 2 l + E_T^{miss}$. The last SR, SR-Zjets,
was designed specifically for $p p \to \bar{\tilde{\chi}}_i^0 \tilde{\chi}_j^\pm \to (\tilde{\chi}_1^0 Z) (\tilde{\chi}_1^0 W^\pm ) \to 2l 2j + E_T^{miss}$.
In our study, we note that the bounds on the chargino mass from the SR-WW are much weaker than those from the SR-Zjets in simplified model \cite{2lETmiss},
and also that the involved simulations are rather time consuming since we have thousands samples,
so we only consider the SR-$m_{T2}$ for direct slepton pair production and the SR-Zjets for chargino and neutralino associated productions to save time.

Both the SR-$m_{T2}$ and the SR-Zjets  require that the signal events contain exactly two same flavor opposite sign (SFOS) leptons with $p_T$ $> $35 GeV and $>$ 20 GeV, and
their invariant mass $m_{ll}$ must be larger than 20 GeV.  Events containing central ($|\eta| <$ 2.4)  b-jets, forward (2.4 $< |\eta| <$ 4.5, $p_T >$ 30 GeV) jets, or $\tau$-jet candidates are rejected. Further selections are applied for the different SRs, which are summarized in Table \ref{table1} with $N_{j}$ representing the number of the central light jets, which are defined as $|\eta| <$ 2.4 and $P_T >$ 20(45) GeV for SR-$m_{T2}$ (SR-Zjets), and $\Delta m_{ll,Z}$ denoting the mass difference between the SFOS lepton pair and the Z boson.
Note that in order to suppress the backgrounds containing two W bosons for the SR-$m_{T2}$,  the 'stransverse' mass $m_{T2}$ is introduced. This quantity is defined by
\begin{eqnarray}
m_{T2}=\mathop{\textrm{min}}\limits_{{\bf q}_T}\left[\textrm{max}(m_T({\bf p}_T^{l1},{\bf q}_T),m_T({\bf p}_T^{l2},{\bf p}_T^{miss}-{\bf q}_T)\right],
\end{eqnarray}
where ${\bf p}_T^{l1}$ and ${\bf p}_T^{l2}$ stand for the transverse momenta of the two leptons, and a varying momentum ${\bf q}_T$ is introduced to minimize the larger one of the two transverse masses $m_T({\bf p}_T,{\bf q}_T)=\sqrt{2(p_Tq_T-{\bf p}_T.{\bf q}_T)}$. By contrast, in order to suppress the background $Z$ + jets production for the SR-Zjets,
the cuts on the transverse momentum $P_T^{ll}$ and the separation angle $\Delta R_{ll} = \sqrt{(\Delta \phi_{ll})^2+(\Delta \eta_{ll})^2}$ of the two leptons, and $E_{T,rel}^{miss}$
are imposed. Here the $E_{T, rel}^{miss}$ is a variant of $E_T^{miss}$, and defined by
\begin{eqnarray}
E_{T,rel}^{miss}=\left\{\begin{array}{ll}
E_T^{miss} 						& {\rm if} \Delta \phi_{l,j} \geqslant \pi/2  \\
E_T^{miss} \times sin\Delta \phi_{l,j}  	& {\rm if} \Delta \phi_{l,j} \leqslant \pi/2
\end{array}\right.,
\end{eqnarray}
where $ \Delta \phi_{l,j}$ is the azimuthal angle between the direction of ${\bf p}_T^{miss}$ and that of the nearest lepton or central jet.

\begin{table}[t]
\small
\caption{The details of the 20 bins defined in the SR0$\tau$a. For each bin the expected cross sections of its SM backgrounds after cuts at the 14 TeV LHC are also presented for later use. All quantities with mass dimension and cross sections are given in units of GeV and fb respectively.}\label{table2}
\centering
\begin{tabular}{c|cccc|ccccccc}
\hline
SR0$\tau$a & $m_{SFOS}$ & $m_T$  & $E_T^{miss}$ & $m_{3l}$ & VVV  & WZ    & ZZ   & t    & h    & $t\bar{t}$ & Total \\
\hline
1          & 12-40      & 0-80   & 50-90        & no       & 0.03 & 1.11  & 0.11 & 0.02 & 0.07 & 1.05       & 2.41  \\
2          & 12-40      & 0-80   & $>$90        & no       & 0.01 & 0.28  & 0.01 & 0.01 & 0.03 & 0.11       & 0.45  \\
3          & 12-40      & $>$80  & 50-75        & no       & 0.02 & 0.66  & 0.03 & 0.00 & 0.07 & 0.22       & 1.00  \\
4          & 12-40      & $>$80  & $>$75        & no       & 0.06 & 0.45  & 0.02 & 0.02 & 0.05 & 0.48       & 1.08  \\
\hline
5          & 40-60      & 0-80   & 50-75        & yes      & 0.02 & 0.52  & 0.13 & 0.01 & 0.04 & 0.65       & 1.37  \\
6          & 40-60      & 0-80   & $>$75        & no       & 0.02 & 0.33  & 0.02 & 0.00 & 0.04 & 0.33       & 0.76  \\
7          & 40-60      & $>$80  & 50-135       & no       & 0.08 & 0.64  & 0.05 & 0.00 & 0.11 & 0.61       & 1.49  \\
8          & 40-60      & $>$80  & $>$135       & no       & 0.02 & 0.04  & 0.00 & 0.02 & 0.04 & 0.08       & 0.20  \\
\hline
9          & 60-81.2    & 0-80   & 50-75        & yes      & 0.02 & 1.40  & 0.12 & 0.02 & 0.03 & 0.79       & 2.40  \\
10         & 60-81.2    & $>$80  & 50-75        & no       & 0.04 & 1.09  & 0.05 & 0.02 & 0.02 & 0.29       & 1.51  \\
11         & 60-81.2    & 0-110  & $>$75        & no       & 0.06 & 1.75  & 0.07 & 0.07 & 0.04 & 0.99       & 2.98  \\
12         & 60-81.2    & $>$110 & $>$75        & no       & 0.07 & 0.34  & 0.01 & 0.02 & 0.02 & 0.16       & 0.63  \\
\hline
13         & 81.2-101.2 & 0-110  & 50-90        & yes      & 0.14 & 52.16 & 2.60 & 0.56 & 0.23 & 10.73      & 66.41 \\
14         & 81.2-101.2 & 0-110  & $>$90        & no       & 0.10 & 19.95 & 0.56 & 0.44 & 0.15 & 0.42       & 21.62 \\
15         & 81.2-101.2 & $>$110 & 50-135       & no       & 0.11 & 5.13  & 0.35 & 0.13 & 0.04 & 0.21       & 5.98  \\
16         & 81.2-101.2 & $>$110 & $>$135       & no       & 0.05 & 0.47  & 0.01 & 0.02 & 0.00 & 0.03       & 0.59  \\
\hline
17         & $>$101.2   & 0-180  & 50-210       & no       & 0.34 & 4.80  & 0.24 & 0.12 & 0.13 & 2.01       & 7.65  \\
18         & $>$101.2   & $>$180 & 50-210       & no       & 0.06 & 0.28  & 0.01 & 0.02 & 0.01 & 0.04       & 0.44  \\
19         & $>$101.2   & 0-120  & $>$210       & no       & 0.02 & 0.13  & 0.00 & 0.00 & 0.00 & 0.08       & 0.24  \\
20         & $>$101.2   & $>$120 & $>$210       & no       & 0.02 & 0.02  & 0.00 & 0.00 & 0.00 & 0.04       & 0.09 \\
\hline
\end{tabular}
\end{table}

\item The search for $3l + E_T^{miss}$ signal from the chargino and neutralino associated production
at the 8 TeV LHC with $20.3 \ {\rm fb}^{-1}$ integrated luminosity \cite{3lETmiss}. Signal events in this analysis were required to contain exactly three leptons and no b-tagged jets. The leptons must be separated from each other by $\Delta R >0.3$, include at least one electron or muon, fire at least one of the single- and double-lepton triggers
and also satisfy the $P_T$-threshold requirements \cite{3lETmiss}. Then according to the flavor and charge of the leptons, five SRs, SR0$\tau$a, SR0$\tau$b, SR1$\tau$, SR2$\tau$a and SR2$\tau$b were defined, and each of them was further designed to detect efficiently a certain type of signal. To be more specific, the SR0$\tau$a was optimized for maximum sensitivity to the chargino and neutralino production followed by the $\tilde{l}_L$-mediated or $W Z$-mediated decay of the sparticles,
the SR0$\tau$b, SR1$\tau$ and SR2$\tau$b are all for the $W h$-mediated decay, and  the SR2$\tau$a targets the $\tilde{\tau}$-mediated decay mode. In the simplified model discussed in \cite{3lETmiss}, the constraint from the SR0$\tau$a is much stronger than that from the SR0$\tau$b, SR1$\tau$ and SR2$\tau$b in limiting the chargino/neutralino sector mainly because the branching ratios of $h$ decays into leptons are small,
and with regard to our scenario, the SR2$\tau$a is less efficient because the branching ratios of $\tilde{\chi}_1^{\pm}\to \tau^\pm \nu_{\tau}\tilde{\chi}_1^0$ and $\tilde{\chi}_i^{0}\to \tau^\pm \tau^\mp \tilde{\chi}_1^0$ are usually small. So in our study, we only consider the SR0$\tau$a for the chargino and neutralino associated production processes.

In the SR0$\tau$a, 20 bins were defined by the invariant mass of the SFOS lepton pair closer to the Z boson mass $m_{SFOS}$, $E_T^{miss}$ and $m_T=\sqrt{2p_T^l E_T^{miss}-2{\bf p}_T^l\cdot{\bf p}_T^{miss}}$ where ${\bf p}_T^l$ is the transverse momentum of the lepton not forming the SFOS lepton pair. The details of the bins are listed in Table \ref{table2}. Note that in bin-5, 9 and 13, events with $|m_{3l}-m_Z|<10$ GeV are vetoed where $m_{3l}$ denotes the trilepton mass.

\end{itemize}

About the considered analyses, we note that the SR-$m_{T2}$ focuses on the slepton pair production process,
and its SRs are statistically dependent since they overlap with each other. So we use the SR of the SR-$m_{T2}$ with the best exclusion limit to determine whether the model point is excluded. We also note that the SRs targeting the neutralino and chargino associated production processes, i.e. SR0$\tau$a and SR-Z jets, are disjoint, which means that their results can be statistically combined to maximize the significance. In our study we combine them together though the $CL_s$ method\cite{CLs} with RooStats\cite{RooStats}, in which the likelihood functions are written as
\begin{eqnarray}\label{fun:CLs1}
\mathcal{L}(n_{i}|s_i+b_i)=\prod\limits_{i=1}^{N_{bin}} \frac{1}{\sqrt{2\pi \sigma_{b_i}^2}}\frac{1}{\sqrt{2\pi \sigma_{s_i}^2}} \int db'_i \int ds'_i \frac{(s'_i+b'_i)^{n_{i}}e^{-(s'_i+b'_i)}}{n_{i}!}e^{\frac{(b_i-b'_i)^2}{2\sigma_{b_i}^2}}e^{\frac{(s_i-s'_i)^2}{2\sigma_{s_i}^2}}
\end{eqnarray}
for signal and
\begin{eqnarray}\label{fun:CLs2}
\mathcal{L}(n_{i}|b_i)=\prod\limits_{i=1}^{N_{bin}} \frac{1}{\sqrt{2\pi \sigma_{b_i}^2}} \int db'_i \frac{b'^{n_i}_i e^{-b'_i} }{n_i!} e^{\frac{(b_i-b'_i)^2}{2\sigma_{b_i}^2}}
\end{eqnarray}
for backgrounds. In above expressions, $n_i$, $s_i$ and $b_i$ are the numbers of observed events, predicted signal events
and background events in each SR or bin respectively, and $\sigma_{s_i}$ and $\sigma_{b_i}$ are the corresponding total systematic uncertainties. In our calculation, we take the values of $n_i$, $b_i$ and $\sigma_{b_i}$ from the experimental reports and fix the relative uncertainties of the signals at $10\%$, i.e. $\sigma_{s_i}/s_i = 10\%$.

In practical calculation, we use MG5$\_$aMC/MadEvents \cite{MadGraph} to generate the tree level events of
the processes contributing to those SRs, and then pass them through PYTHIA\cite{PYTHIA} for parton showering and hadronization and DELPHES\cite{DELPHES} for fast simulation of the ATLAS detector. The SRs described above have been implemented by CheckMATE\cite{CheckMATE}, and the involved cross sections are calculated by the code PROSPINO2\cite{PROSPINO}\footnote{About this point we emphasize that we simulate all six neutralino and chargino production processes contributing to the SRs, i.e. $p p \to \tilde{\chi}_i^0 \tilde{\chi}_j^\pm$ with $i=2,3,4$ and $j=1,2$, and add their contributions into the signal events
for every SR (bin).}. After these procedures, we can determine whether the model points survive the constraints from the direct searches.

The results of the direct searches at the $8 \ {\rm TeV}$ LHC are showed in Fig.\ref{fig1}, where the blue points are excluded at $95\%$ C.L. by the SR-$m_{T2}$, the green ones are excluded by the combination of the SR0${\tau a}$ and  the SR-Zjets, and the red and orange ones are the remaining samples with the former able to get the measured DM relic
density at $3 \sigma $ level and the latter only satisfying the $3 \sigma$ upper bound of the density. From Fig.\ref{fig1},
one can learn the following facts:
\begin{itemize}
\item After considering the constraints from the SR-$m_{T2}$, most Type-I samples are excluded, especially for those with $m_{\tilde{\chi}_1^0} \leq 50$ GeV. In this case, the sleptons are usually heavier than about 250 {\rm GeV}
    for $m_{\tilde{\chi}_1^0} \simeq m_Z/2$ and $m_{\tilde{\chi}_1^0} \simeq m_h/2$, and $\mu$ is less than $470 {\rm GeV}$ and $680 {\rm GeV}$ for $m_{\tilde{\chi}_1^0} \simeq m_Z/2$ and $m_{\tilde{\chi}_1^0} \simeq m_h/2$ respectively.
\item The combination of the SR0${\tau a}$ and the SR-Zjets can only exclude the samples with $\mu \lesssim 220$ GeV, which is much weaker than the exclusion limit for the chargino mass reported in \cite{2lETmiss} and \cite{3lETmiss}. One reason is that the lighter chargino in our scenario is higgsino-dominated instead of wino-dominated.
    As a result, the neutralino and chargino associated production rate is relatively small. Another reason is that in our scenario the higgsino-dominated $\tilde{\chi}_2^0$ and $\tilde{\chi}_3^0 $ may decay into $Z \tilde{\chi}_1^0$, $h \tilde{\chi}_1^0$ and $\tilde{l}^\ast l$, and the higgsino-dominated $\tilde{\chi}_1^\pm$ may decay into $W \tilde{\chi}_1^0$ and $\tilde{l} \nu$. Consequently, the trilepton signal is suppressed.
\item There exist samples with $\mu \sim 150 {\rm GeV}$ on the right bottom of the $m_Z/2$ peak and the
      left bottom of the $m_h/2$ peak in Fig.\ref{fig1} which can not be excluded by the combination of the SR0${\tau a}$ and the SR-Zjets. There also exist some samples with $m_{\tilde{l}} \lesssim 180 {\rm GeV}$ for $m_{\tilde{\chi}_1^0}$ changing from $70 {\rm GeV}$ to $100 {\rm GeV}$ which can not be excluded by the SR-$m_{T2}$. All these samples are characterized by the compressed spectrum of the parent sparticle with respect to their decay products. In such a situation, the acceptance efficiencies of the signals in our discussion are rather low.
\end{itemize}

At this stage, we'd like to clarify the differences of our study from previous literatures \cite{LDM-23} and \cite{LDM-27}. In \cite{LDM-23}, the authors scanned the parameter space of the MSSM by relaxing the slepton masses
to get the light DM scenario, which is quite similar to what we do in this work. The main difference of the two works
is that the authors of \cite{LDM-23} used the package SModelS \cite{SmodelS} to consider the constraints of the direct searches
on the electroweak-inos and sleptons, while we do it by detailed simulations. Because the feasibility of the SModelS is based on certain assumptions (e.g. the approximate degeneracy
of $\tilde{\chi}_1^\pm$ and $\tilde{\chi}_2^0$) which can not be applied to some of our samples, and also because it considers separately the signals coming
from different sparticles that lead to the same final state \cite{SmodelS}, the constraints of the SModelS on the electroweak-inos
should be conservative. In fact, we once compared the difference of the two methods in implementing the constraints, and
verified this conclusion. In \cite{LDM-27}, the authors got the light DM scenario by decoupling all sparicles except for the bino-like $\tilde{\chi}_1^0$ and the higgsino-like $\tilde{\chi}_2^0$, $\tilde{\chi}_3^0$ and $\tilde{\chi}_1^\pm$. The advantage of such a simplification is that, without the participation of light sleptons, the correlation of $m_{\tilde{\chi}_1^0}$ with $\mu$ is rather clear, but as we have shown in this work, the sleptons not only played an important role in the $\tilde{\chi}_1^0$ annihilation, but also affect the decays of the electroweak-inos. So the impact of the light sleptons on the scenario should be taken into account. Another difference of \cite{LDM-27} from our work is that the work \cite{LDM-27} only considered the trilepton signal to limit the light DM scenario, while we combine the dilepton and trilepton signals to limit the scenario.

\section{Test the light DM scenario at the 14 TeV LHC}

\begin{figure}[t]
  \centering
  \includegraphics[width=15cm]{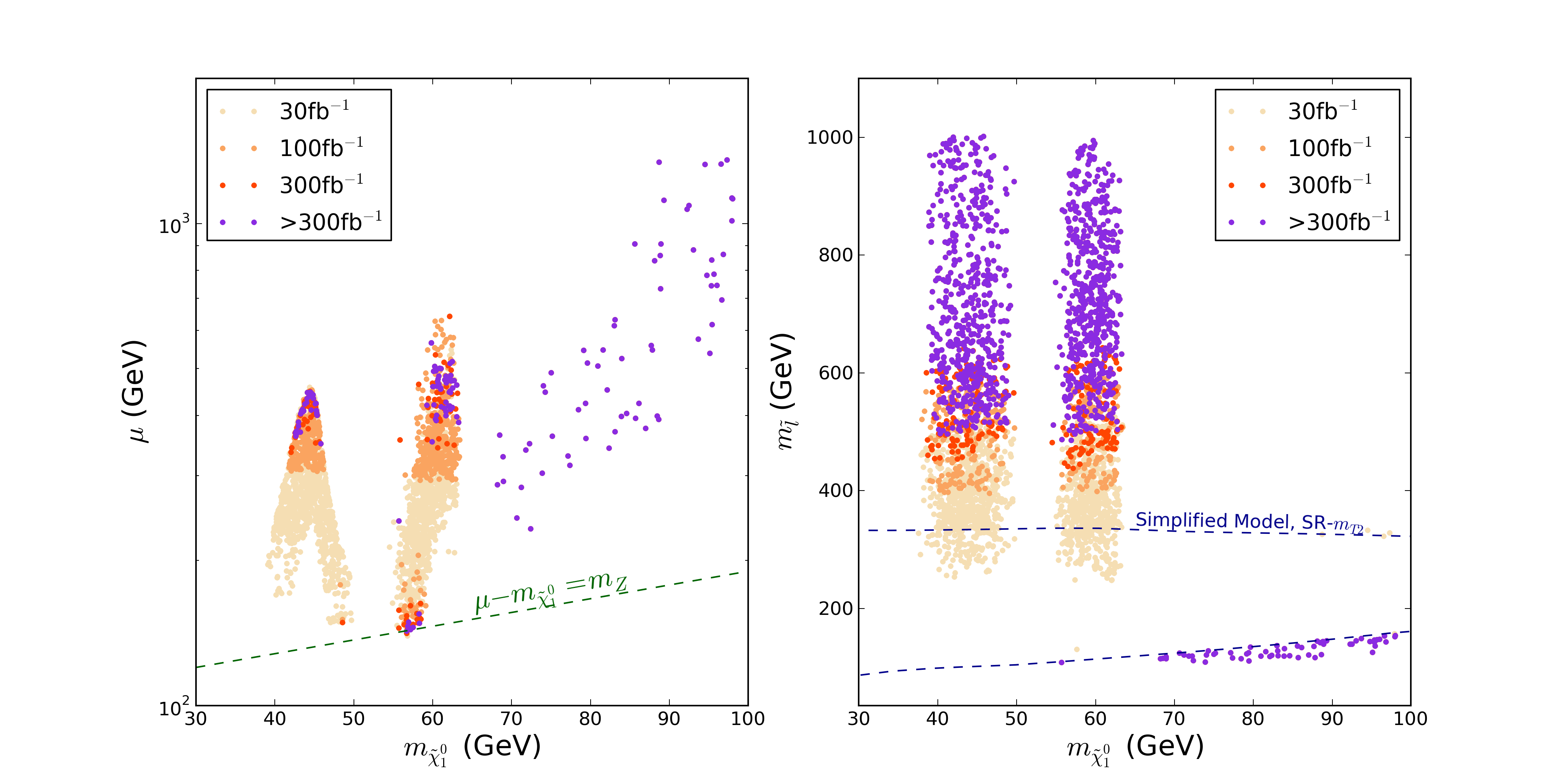}
  \caption{The 95\% exclusion bounds of the 14 TeV LHC by the combination of the SR0${\tau a}$ and the SR-Zjets and by the SR-$m_{T2}$, which are projected on the $m_{\tilde{\chi}_1^0}-\mu$
  plane (left panel) and the $m_{\tilde{\chi}_1^0}-m_{\tilde{l}}$ plane (right panel) respectively for the samples surviving the constraints considered in sections 3 and 4. The samples marked by the colors faint yellow, brown and orange will be excluded at 95\% C.L. by the integrated luminosities of 30 $fb^{-1}$, 100 $fb^{-1}$ and 300 $fb^{-1}$ respectively,  and those marked by the color violet can not be excluded even with 300 $fb^{-1}$ integrated data.}
  \label{fig2}
\end{figure}

From the discussion in last section, one can learn that the searches for the sleptons and the electroweak-inos at the 8 TeV LHC have important impact on the light DM scenario, e.g. lots of the samples of the scenario have been excluded. Given the ongoing of the upgraded LHC, one may expect that much tighter constraints on the scenario will be obtained, and even some sparticles in this scenario will be discovered in near future.

We investigate this issue by considering the slepton pair production and the neutralino and chargino associated
production at the 14 TeV LHC. For simplicity, we assume the same cuts at the 14 TeV LHC as those in the SR-$m_{T2}$, the SR0${\tau a}$
and the SR-Zjets at the 8 TeV LHC, and get the SM backgrounds of the signals by two steps. We first simulate
each background process at the 8 TeV LHC, and compare the simulated event number in each SR with its validated number, which
was obtained by experimentalists, to get a correction factor (this factor usually varies from 1 to 5 from our simulation). Subsequently we suppose that the dominant backgrounds at the 14 TeV LHC come from
the same processes as those at the 8 TeV LHC, which include $WW$, $ZV$, $Z$ + jets and
top quark production for $2l+E_T^{miss}$, and diboson, $t\bar{t}V$, $tZ$, $VVV$ and Higgs boson
production for $3l+E_T^{miss}$, and simulate each of them at the 14 TeV LHC. Then we take the simulation results for the 14 TeV LHC
multiplied with the corresponding correction factors as our predictions of the backgrounds,
which are given in Table \ref{table1} and Table \ref{table2}. We realize that the backgrounds obtained in this way only act as rough estimates
of the true backgrounds at the time when we have no detailed information about the ATLAS detector at the 14 TeV LHC.

In Fig.\ref{fig2}, we show our simulation results for the direct production of the charginos and neutralinos at the
14 TeV LHC on the $m_{\tilde{\chi}_1^0}-\mu$ plane (the left panel) and those for the direct production
of the sleptons on the $m_{\tilde{\chi}_1^0}-m_{\tilde{l}}$ plane (the right panel). The samples we considered in this figure
are those surviving all the constraints considered in sections 3 and 4.
The exclusion significance is calculated by the CLs method with the $n_i$ in Eq.(\ref{fun:CLs1}) and
Eq.(\ref{fun:CLs2}) set to be $b_i$ and the total relative systematic uncertainties of the backgrounds and the signals taken
same as those at the 8 TeV LHC. The samples marked by the colors faint yellow, brown and orange are those
which will be excluded at 95\% C.L. with the integrated luminosity of 30 $fb^{-1}$, 100 $fb^{-1}$ and 300 $fb^{-1}$
respectively, and those marked by the color violet denote the samples that can not be excluded with
300 $fb^{-1}$ data.

From the left panel of Fig.\ref{fig2}, one can clearly see an exclusion line at $\mu \simeq 300GeV$ for $30fb^{-1}$
integrated luminosity, and the limit raises with the increase of the luminosity. This means that, except for
the compressed spectrum case, a stricter bound on $\mu$ can be set by the 14 TeV LHC if no excess of the multi-lepton signals
is observed. Moreover, we note the existence of a few samples on the top of the $m_Z/2$ peak which are hard to be excluded
by the direct search for the electroweak-inos even for $300 {\rm fb^{-1}}$ integrated luminosity.
For these samples, we checked that they are characterized by predicting relatively light sleptons, which can act as the decay product
of the electroweak-inos.  From the right panel of Fig.\ref{fig2}, one can learn that except for the
samples with the compressed spectrum, lower bounds of about $400 {\rm GeV}$, $450 {\rm GeV}$ and
$500 {\rm GeV}$ on sleptons masses can be obtained with the integrated luminosities of 30 $fb^{-1}$, 100 $fb^{-1}$ and
300 $fb^{-1}$ respectively. Especially, we note that for the
300 $fb^{-1}$ luminosity case, the sleptons must be heavier than the higgsino-like $\tilde{\chi}_2^0$ and $\tilde{\chi}_3^0$
predicted by the samples around the $m_Z/2$ peak, which means that
the neutralinos around the $m_Z/2$ peak can not decay into the on-shell sleptons any more.

\begin{figure}[t]
\centering
\includegraphics[width=15cm]{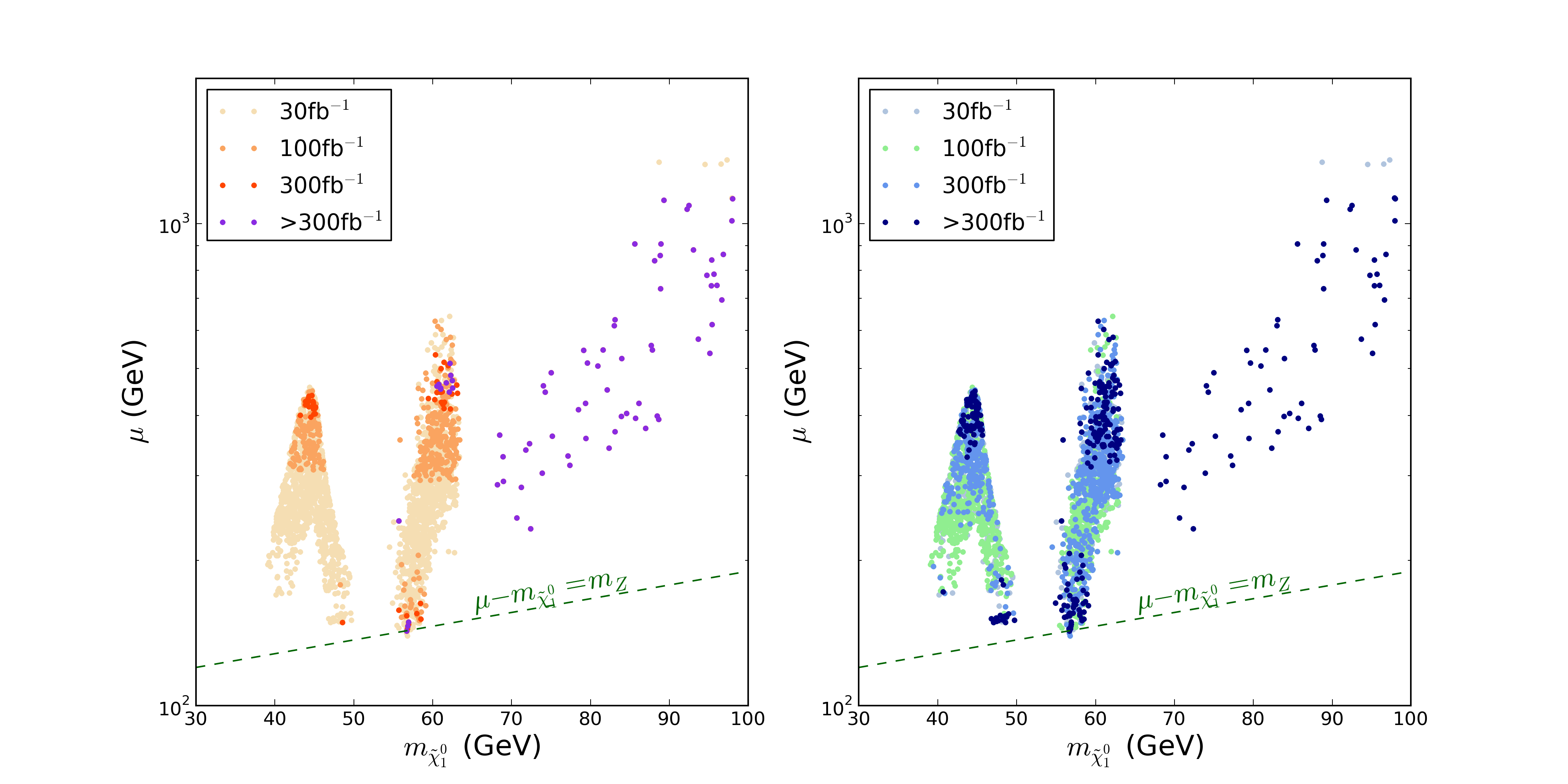}
\caption{The 95\% exclusion capability (left panel) and the discovery capability (right panel) of the multi-lepton searches at the 14 TeV LHC,
which are projected on the $m_{\tilde{\chi}_1^0}-\mu$ plane for the samples surviving the constraints considered in sections 3 and 4.
The color convention in the left panel is same as that in Fig.\ref{fig2}. The colors light blue, green and blue in the right panel
denote the samples which can be discovered at $ 5 \sigma$ C.L. with the data of 30 $fb^{-1}$, 100 $fb^{-1}$ and 300 $fb^{-1}$ respectively,
and the color black represents the difficult case of the light DM scenario, namely the samples that fail to be discovered even with 300 $fb^{-1}$ luminosity.}
\label{fig3}
\end{figure}

Next we consider the search for the sleptons and that for the electroweak-inos simultaneously. In Fig.\ref{fig3}, we again focus on the samples surviving the constraints listed in sections 3 and 4, and project the $95\%$ exclusion capability and the discovery capability of the searches on the $m_{\tilde{\chi}_1^0}-\mu$ plane, which are shown in the left panel and right panel of Fig.\ref{fig3} respectively. Here the discovery significance is also calculated by the CLs method, but with the $n_i$ in Eq.(\ref{fun:CLs1}) and Eq.(\ref{fun:CLs2}) taken to be $b_i + s_i$.
The colors in the left panel have same meanings as those in Fig.\ref{fig2}, while the colors light blue, green and blue in the right panel denote the samples which can be discovered at
$ 5 \sigma$ C.L. with the data of 30 $fb^{-1}$, 100 $fb^{-1}$ and 300 $fb^{-1}$ respectively. Note the color black in the right panel represents the difficult cases of the light DM scenario, namely
the samples that fail to be discovered even with 300 $fb^{-1}$ luminosity. From Fig.\ref{fig3}, one can learn following facts:
\begin{itemize}
\item In case that no multi-lepton plus $E_T^{miss}$ signals are observed at the 14 TeV LHC, a lower limit on $m_{\tilde{\chi}_1^0}$ will be set at 42 GeV and 44 GeV with 30 $fb^{-1}$ and  100 $fb^{-1}$ data respectively, and the limit will be further pushed up to $55 \ {\rm GeV}$  with 300 $fb^{-1}$ data.
\item If the light DM scenario is chosen by nature, and meanwhile no multi-lepton plus $E_T^{miss}$ signals are observed at the 14 TeV LHC with  300 $fb^{-1}$ data, most samples of the light DM scenario will be excluded. In this case, $\tilde{\chi}_1^0$ must annihilate in early universe either by $s$-channel exchange of the SM-like Higgs boson or by $t/u$-channel slepton mediation. The latter situation then requires that the sleptons must be lighter than about $170 {\rm GeV}$, and their mass splitting from the $\tilde{\chi}_1^0$ should be moderately small.
\item At the 14 TeV LHC with 300 $fb^{-1}$ data, a large portion of the samples in the light DM scenario can be discovered. These samples are characterized by $\mu \lesssim 350 {\rm GeV}$ and a sizable splitting between $\mu$ and $m_{\tilde{\chi}_1^0}$.
\item We note from the right panel that there also exist samples with $\mu \simeq 700 {\rm GeV}$ at the $m_h/2$ peak which will be discovered at the 14 TeV LHC. For these samples, the wino mass $M_2$ is smaller than $600 {\rm GeV}$, so it is actually wino-dominated neutralino and chargino that contribute dominantly to the multi-lepton signals.
\end{itemize}

From above discussions, one can learn that the light sleptons may play an important role in the light DM scenario. Here we emphasize that our conclusions are based on a
common slepton mass assumption, and if only $\tilde{\tau}$s are assumed to be light, the conclusions may change slightly. Indeed, for the latter case a study similar to what we do must be done, and new experiment pertinent to the $\tilde{\tau}$ search should be considered \cite{LDM-20}.  We also want to emphasize that the cuts in our simulation at
the 14 TeV LHC can be optimized, and meanwhile the pileup effect should be estimated. This is beyond the scope of this work.

\section{Future DM direct searches}

\begin{figure}[t]
  \centering
  \includegraphics[width=12cm]{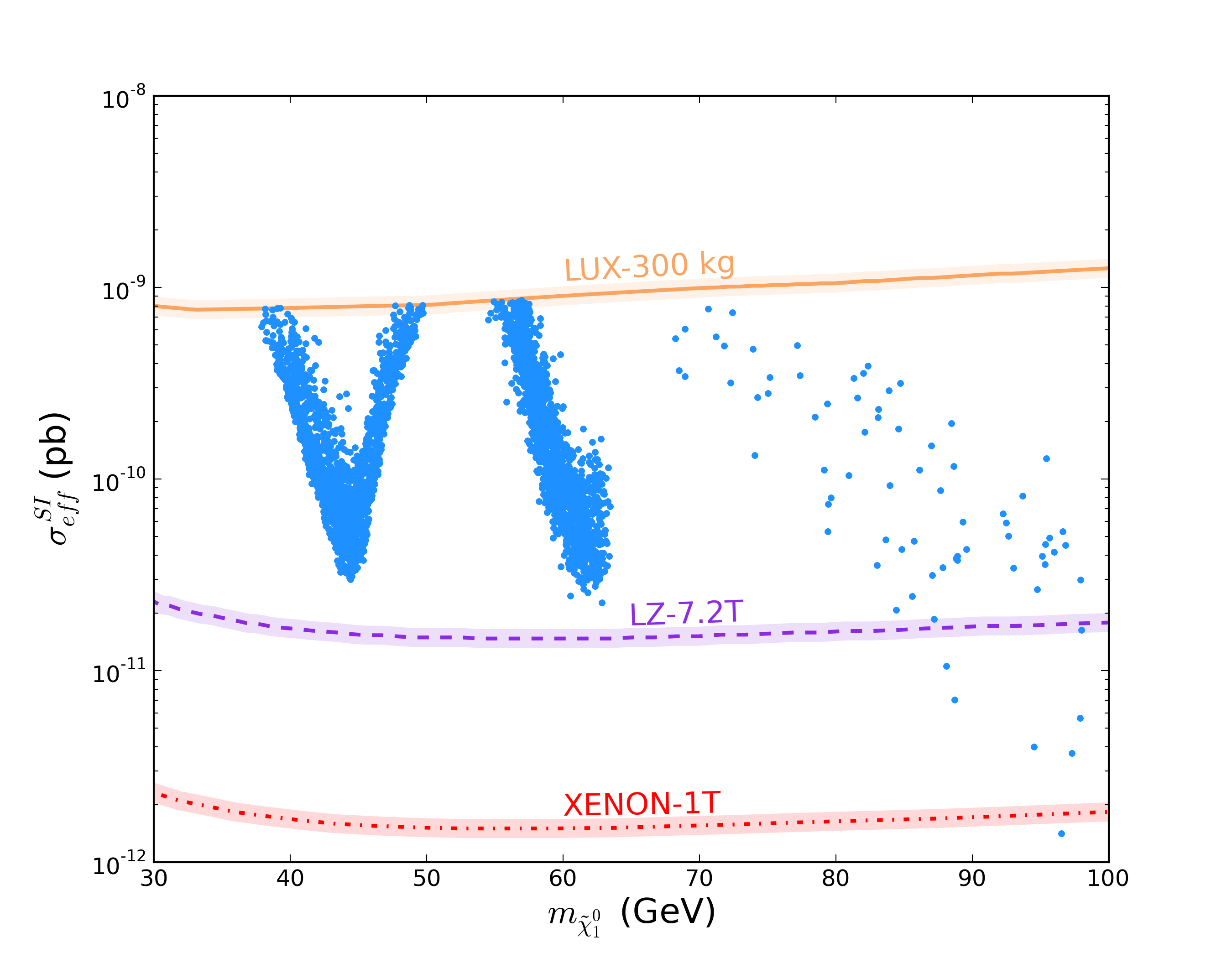}
  \caption{The spin independent DM-nucleon scattering cross section versus the DM mass for the samples surviving the constraints in section 3 and section 4. The capabilities of future DM direct detection experiments in detecting the cross section are also plotted.}
  \label{fig4}
\end{figure}

As a supplement to the discussion in section 5, we investigate the capabilities of the future DM direct search experiments in exploring the light DM scenario.
For this end, we focus on the effective spin independent (SI) DM-nucleon scattering cross section $\sigma^{SI}_{eff}$, which is defined by $\sigma^{SI}_{eff} = \epsilon \times \sigma^{SI}_{\tilde{\chi}_1^0 p}$ with $\epsilon$ being the fraction of the $\tilde{\chi}_1^0$ in total DM and $\sigma^{SI}_{\tilde{\chi}_1^0 p}$ being the SI $\tilde{\chi}_1^0-p$
scattering rate, and calculate it by the package micrOMEGAs\cite{micrOMEGA} with its default setting $\sigma_{\pi N} = 34  {\rm MeV}$ and $\sigma_0 = 42  {\rm MeV}$
\footnote{We note that if we take $\sigma_{\pi N} = 59  {\rm MeV}$ from \cite{SI-piN} and $\sigma_0 = 58  {\rm MeV}$ from \cite{SI-pi0},
the SI cross section will be enhanced by a factor from $20\%$ to $40\%$.}. In Fig.\ref{fig4},
we display $\sigma^{SI}_{eff}$ versus $m_{\tilde{\chi}_1^0}$ for the samples surviving the constraints in section 3 and section 4 together with the detection limits of future
underground DM direct searches LZ-7.2T and XENON-1T\cite{direct:future}. This figure indicates that most samples of the light DM scenario, especially all the Type-II
and Type-III samples, will be tested by the experiment LZ-7.2T. This figure also indicates that $\sigma^{SI}_{eff}$ dips at $m_{\tilde{\chi}_1^0} \simeq m_Z/2$
and $m_{\tilde{\chi}_1^0} \simeq m_h/2$.  This behavior can be understood by following formulas \cite{Cao-MSSM}:
\begin{eqnarray}
\sigma^{SI}_{eff} &\propto& \biggl( \frac{C_{h \tilde{\chi}_1^0 \tilde{\chi}_1^0}  C_{hqq}}{m_{h}^2}
    + \frac{C_{H  \tilde{\chi}_1^0 \tilde{\chi}_1^0}  C_{Hqq}}{m_{H}^2}     \biggr)^2,  \nonumber \\
C_{h \tilde{\chi}_1^0 \tilde{\chi}_1^0} &\simeq&
  \frac{m_Z \sin \theta_W \tan \theta_W}{M_1^2 - \mu^2}
    \bigl[ M_1 + \mu \sin2 \beta \bigr], \nonumber \\
C_{H \tilde{\chi}_1^0 \tilde{\chi}_1^0} &\simeq&
   - \frac{m_Z \sin \theta_W \tan \theta_W}{M_1^2 - \mu^2}
    \mu \cos2\beta,  \label{SI-Cross}
\end{eqnarray}
where $C_{XYZ}$ stands for the Yukawa couplings of the CP-even Higgs bosons $h$ and $H$, and the fact that for the case
$m_{\tilde{\chi}_1^0} \simeq m_Z/2$ or $m_{\tilde{\chi}_1^0} \simeq m_h/2$, the value of $\mu$ tends to be large.  From Eq.(\ref{SI-Cross}) one can also infer that
in the case of a large $\tan \beta$, the $H$-meidated contribution may still be significant even for a heavy $H$ because $C_{Hqq} $ for down-type quarks is proportional to $\tan \beta$ and $|\cos 2 \beta| \simeq 1$. Anyhow, $\sigma^{SI}_{eff}$ usually decreases as $H$ becomes heavier.

\section{Conclusion}

In past several years, fruitful results in the searches for sparticles have been obtained at the Run-I of the LHC, which set stronger limits on the spectrum of the sparticles than the LEP experiments.
Now with the operation of the upgraded LHC, it is widely expected that much heavier sparticles will be tested in near future. Obviously, discussing the potential of the LHC experiments to test the MSSM is an important task for both theorists and experimentalists.

In this work we investigate the impact of the sparticle searches at the LHC on the light DM scenario of the MSSM, for
which the DM relic density has put non-trivial constraints on the sparticle spectrum.
We start our study by scanning the vast parameter space of the MSSM to get the samples of the scenario. During the scan, we have considered some easily available constraints, such as those from the DM relic density, the LUX experiment, the searches for the Higgs bosons at colliders as well as B-physics. Next we pay special attention to the important constraints from the direct
searches for the sparticles at the 8 TeV LHC, and investigate how and to what extent the samples are limited. For this end, we simulate the $2l+E_T^{miss}$ signal from slepton
pair production process and the  $2l+E_T^{miss}$ and $3l+E_T^{miss}$ signals from chargino and neutralino associated production processes, and we find that the 8 TeV LHC has
excluded a sizable portion of the samples. Subsequently we extend the simulation study to the 14 TeV LHC and conclude that
the 14 TeV LHC is much more powerful than the 8 TeV LHC in testing the scenario. Explicitly speaking, we obtain following conclusions
\begin{itemize}
\item In case that no multi-lepton plus $E_T^{miss}$ signals are observed at the 14 TeV LHC, a lower limit on $m_{\tilde{\chi}_1^0}$ will be set at 42 GeV and 44 GeV with 30 $fb^{-1}$ and  100 $fb^{-1}$ data respectively, and the limit will be further pushed up to $55 \ {\rm GeV}$  with 300 $fb^{-1}$ data.
\item If the light DM scenario is chosen by nature, and meanwhile no multi-lepton plus $E_T^{miss}$ signals are observed at the 14 TeV LHC with  300 $fb^{-1}$ data, most samples of the light DM scenario will be excluded. In this case, $\tilde{\chi}_1^0$ must annihilate in early universe either by $s$-channel exchange of the SM-like Higgs boson or by $t/u$-channel slepton mediation. The latter situation then requires that the sleptons must be lighter than about $170 {\rm GeV}$, and their mass splitting from the $\tilde{\chi}_1^0$ should be moderately small.
\item At the 14 TeV LHC with 300 $fb^{-1}$ data, a large portion of the samples in the light DM scenario can be discovered. These samples are characterized by $\mu \lesssim 350 {\rm GeV}$ and a sizable splitting between $\mu$ and $m_{\tilde{\chi}_1^0}$.
\end{itemize}
At the end of our work, we also discuss the capability of the future DM direct detection experiments to test the scenario. We conclude that, for the parameter space we considered, most samples of the scenario can be covered by the LUX-ZEPLIN 7.2 Ton experiment.

{\bf Note added}: At the final stage of this work, the paper \cite{Hamaguchi:2015rxa} appeared, which also studied the impacts of the direct search for the multi-lepton signals at the 14 TeV LHC
on the light DM scenario in the MSSM. Although we adopt different SRs from those in \cite{Hamaguchi:2015rxa} in the search, we get same conclusion that
 the samples with $\mu \lesssim 500 {\rm GeV}$ in the light DM scenario will be excluded  with 300 $fb^{-1}$ integrated luminosity data. The main difference of the two works
 is that in \cite{Hamaguchi:2015rxa}, the authors got the light DM scenario by fixing $\Omega_{\tilde{\chi}_1^0} h^2 \simeq 0.120$ and varying $M_1$ and $\mu$, while we get the scenario by an intensive scan over the vast parameter space of the MSSM.

\end{document}